\documentclass[12pt]{article}
\usepackage{amssymb}

\begin{document}

\title{Integration in General Relativity}
\author{\small Andrew DeBenedictis}
\date{Dec. 03, 1995}
\maketitle 

\begin{abstract}
This paper presents a brief but comprehensive introduction to certain 
mathematical techniques in General Relativity. Familiar mathematical 
procedures are investigated taking into account the complications of 
introducing a non trivial space-time geometry. This transcript should be 
of use to the beginning student and assumes only very basic familiarity 
with tensor analysis and modern notation. This paper will also be of use 
to gravitational physicists as a quick reference.
\end{abstract}
\vspace{0.5cm} 
\noindent\textbf{Conventions}\newline
The following notation is used: The metric tensor, $g_{\mu\nu}$, has a 
signature of +2 and $g=\left|det\left(g_{\mu\nu}\right)\right|$. Semi-colons 
denote covariant derivatives while commas represent ordinary derivatives.

\section{Introduction}
\qquad Say we have a tensor $\mathbf{T}$ then, like the partial derivative,
the covariant derivative can be thought of as a limiting value of a
difference quotient. A complication arises from the fact that a tensor at
two different points in space-time transform differently. That is, the
tensor at point $r$ will transform in a different way than a tensor at point
$r+dr$. In polar coordinates we are familiar with the fact that the 
length and
direction of the basis vectors change as one goes from one point to another.
If we have a vector written in terms of its basis as follows, 
\begin{equation}
\mathbf{V}=V^\alpha \mathbf{e}_\alpha
\end{equation}
the derivative with respect to co-ordinate $x^\beta$ would be: 
\begin{equation}
\frac{\partial V^\alpha }{\partial x^\beta }\mathbf{e}_\alpha +V^\alpha 
\frac{\partial \mathbf{e}_\alpha }{\partial x^\beta }.
\end{equation}
We define the \textit{Christoffel symbol} $\Gamma _{\alpha \beta }^\mu $ as
representing the coefficient of the $\mu ^{th}$ component of $\frac{\partial 
\mathbf{e}_\alpha }{\partial x^\beta }$. The above derivative becomes (after
re\-labelling dummy indices),
\begin{equation}
\left( V^\alpha ,_\beta +V^\mu \Gamma _{\mu \beta }^\alpha \right) \mathbf{e}%
_\alpha .
\end{equation}

When it comes to integration, we are performing the reversal of a partial
differentiation and can therefore not just integrate a covariant derivative.
Also, an integral over tensor components does not give a result which is a
tensor whereas integration over a scalar does.

We can convert expressions such as $P_{;\nu }^\nu $ into an expression
containing only partial derivatives as follows: First write out the
expression $P_{;\nu }^\nu $ in terms of the Christoffel symbol 
\begin{equation}
P_{;\nu }^\nu =P_{,\nu }^\nu +P^\lambda \Gamma _{\lambda \nu }^\nu .
\end{equation}
Now use the fact that 
\begin{eqnarray}
\Gamma _{\lambda\nu }^{\nu} &=&\frac {1}{2}g^{\nu \alpha }\left( g_{\nu\alpha
,\lambda}+g_{\alpha \lambda ,\nu }-g_{\lambda \nu ,\alpha }\right) \\
&=&\frac {1}{2}g^{\nu \alpha }\left( g_{\alpha \lambda ,\nu }-g_{\lambda \nu
,\alpha }\right) +\frac {1}{2}g^{\nu \alpha }g_{\nu \alpha ,\lambda }.  \nonumber
\end{eqnarray}
The first term in the last expression is equal to zero since it is $g^{\nu
\alpha }$ multiplied by a tensor which is anti\-sym\-metric in $\nu 
,\alpha $. Therefore: 
\begin{equation}
\Gamma _{\lambda \nu }^\nu =\frac {1}{2}g^{\nu \alpha }g_{\nu \alpha ,\lambda }.
\end{equation}
Using the fact that $g_{,\lambda }=gg^{\nu \alpha }g_{\alpha \nu ,\lambda }$
gives, 
\begin{eqnarray}
\Gamma _{\lambda \nu }^\nu &=&\frac {1}{2}g_{,\lambda }g^{-1} \\
&=&\frac{\left( \sqrt{g}\right) _{,\lambda }}{\left( \sqrt{g}\right) }%
=\left( \ln \sqrt{g}\right) _{,\lambda }.  \nonumber
\end{eqnarray}
We can now write 
\begin{eqnarray}
P_{;\nu }^\nu &=&P_{,\nu }^\nu +P^\lambda \frac{\left( \sqrt{g}\right)
_{,\lambda }}{\left( \sqrt{g}\right) } \\
&=&\frac 1{\sqrt{g}}\left( \sqrt{g}P^\nu \right) _{,\nu }.\mbox{(after
relabelling dummy indicies)}  \nonumber
\end{eqnarray}

\medskip This result is useful because it allows us to apply Gauss' law
which we know applies to partial derivatives. Gauss' law states that the
volume integral of a divergence can be re-written as an integral over the
boundary surface as follows: 
\begin{equation}
\int P_{,\alpha }^\alpha dV=\oint P^\alpha \widehat{n}_\alpha dS.
\end{equation}
Where $\widehat{n}_\alpha $ is the outward unit normal to the surface. In
our case we need to integrate over proper volume and therefore must use
proper surface area whose element is $\sqrt{g}d^3S$. Therefore, in general
relativity, Gauss' law is generalized to 
\begin{equation}
\int P_{;\alpha }^\alpha \sqrt{g}d^{4}x=\int \left( \sqrt{g}P^\nu \right)
_{,\nu }d^{4}x=\oint P^{\nu} \widehat{n}_{\nu} \sqrt{g}d^{3}x.\medskip
\end{equation}
\subsection{Tensor Densities}
Ordinary tensors transform according to the following transformation law: 
\begin{equation}
T_\nu ^{\prime \mu }=\frac{\partial x^{\prime \mu }}{\partial x^\alpha }%
\frac{\partial x^\beta }{\partial x^{\prime \nu }}T_\beta ^\alpha.
\label{eq:transf}
\end{equation}
An object which transforms according to (\ref{eq:transf}) is called a tensor 
density of
weight zero. A tensor density $\mathfrak{\Im }$ of weight $w$ transforms as
follows: 
\begin{equation}
\mathfrak{\Im }_\nu ^{\prime \mu }=\left| \frac{\partial x}{\partial 
x^{\prime }}%
\right| ^w\frac{\partial x^{\prime \mu }}{\partial x^\alpha }\frac{\partial
x^\beta }{\partial x^{\prime \nu }}\mathfrak{\Im }_\beta ^\alpha,
\end{equation}
which is similar to (\ref{eq:transf}) except for the Jacobian term raised to 
the power of
$w$. We can convert tensor densities of weight $w$ to ordinary tensors by 
noting the transformation of the metric's determinant. 
\begin{eqnarray}
g^{\prime } &=&\left| g_{\gamma ^{\prime }\kappa ^{\prime }}\right| =\left|
g_{\alpha \beta }A_{\gamma ^{\prime }}^\alpha A_{\kappa ^{\prime }}^\beta
\right| \\
&=&\left| g_{\alpha \beta }\right| \left| \frac{\partial x^\alpha }{\partial
x^{\prime \gamma }}\right| \left| \frac{\partial x^\beta }{\partial
x^{\prime \kappa }}\right|  \nonumber \\
&=&\left| \frac{\partial x}{\partial x^{\prime }}\right| ^2g.  \nonumber
\end{eqnarray}
Therefore we can write 
\begin{equation}
\left( g^{\prime }\right) ^{-w/2}\Im _\nu ^{\prime \mu }=\left| \frac{%
\partial x}{\partial x^{\prime }}\right| ^{-w}\left| \frac{\partial x}{%
\partial x^{\prime }}\right| ^w\frac{\partial x^{\prime \mu }}{\partial
x^\alpha }\frac{\partial x^\beta }{\partial x^{\prime \nu }}\Im _\beta
^\alpha
\end{equation}
which transforms like an ordinary tensor (i.e. tensor density of weight
zero). It is these types of tensor densities which we want to consider when
integrating. For example, consider the volume element $d^4x^{\prime }=\left| 
\frac{\partial x^{\prime }}{\partial x}\right| d^4x$. The corresponding
invariant volume element (the proper element) is $\sqrt{g^{\prime }}d^4x=%
\sqrt{g}d^4x.$ We see that $d^4x$ has a weight of -1 since $\sqrt{g}$ has
a weight of +1.
\newline
\newline
\textit{\textbf{The covariant derivative of a scalar density of arbitrary 
weight}}%
\medskip 

The scalar field of weight $w$ transforms as 
\begin{equation}
\Phi^{\prime }=\left| \frac{\partial x}{\partial x^{\prime }}\right| ^w\Phi
.
\end{equation}
Taking the derivative of this creature we get 
\begin{equation}
\frac{\partial \Phi ^{\prime }}{\partial x^{^{\prime }\iota }}=\left| \frac{%
\partial x}{\partial x^{\prime }}\right| ^w\frac{\partial \Phi }{\partial
x^\alpha }\frac{\partial x^\alpha }{\partial x^{^{\prime }\iota }}+w\left| 
\frac{\partial x}{\partial x^{\prime }}\right| ^w\frac{\partial x^{^{\prime
}\alpha }}{\partial x^\beta }\frac{\partial ^2x^\beta }{\partial x^{^{\prime
}\iota }\partial x^{^{\prime }\alpha }}\Phi .
\end{equation}
Noting that the transformation property of the Christoffel symbol is 
\begin{equation}
\Gamma _{\alpha \iota }^{^{\prime }\alpha }=\Gamma _{\sigma \alpha }^\sigma 
\frac{\partial x^\alpha }{\partial x^{^{\prime }\iota }}+\frac{\partial
^2x^\sigma }{\partial x^{^{\prime }\alpha }\partial x^{^{\prime }\iota }}%
\frac{\partial x^{^{\prime }\alpha }}{\partial x^\sigma }.
\end{equation}
This equation can be multiplied by $w\Phi ^{\prime }$ and subtracted from
the previous one to get 
\begin{equation}
\frac{\partial \Phi ^{\prime }}{\partial x^{^{\prime }\iota }}-w\Phi
^{\prime }\Gamma _{\alpha \iota }^{^{\prime }\alpha }=\left| \frac{\partial x%
}{\partial x^{\prime }}\right| ^w\left( \frac{\partial \Phi }{\partial
x^\alpha }-w\Phi \Gamma _{\sigma \alpha }^\sigma \right) \frac{\partial
x^\alpha }{\partial x^{^{\prime }\iota }}
\end{equation}
which displays the transformation properties of $\Phi$ and is its covariant
deriv\-ative.$\blacksquare \medskip $

The above result can be used to find the covariant derivative of a tensor
density of arbitrary weight. Let $\wp ^\mu $ be a contravariant tensor of
weight w which we want to take the covariant derivative of. This can be
written as follows, 
\begin{eqnarray}
\wp _{;\rho }^\mu  &=&\left( \sqrt{g}^w\sqrt{g}^{-w}\wp ^\mu \right)
_{;\rho }  \nonumber \\
&=&\left( \sqrt{g}^w\right) _{;\rho }\sqrt{g}^{-w}\wp ^\mu +\sqrt{g}%
^w\left( \sqrt{g}^{-w}\wp ^\mu \right) _{;\rho }.
\end{eqnarray}
The first term in the last expression is equal to zero from (18) and noting
(7). The second term is a covariant derivative of a tensor of weight zero
multiplied by the factor $\sqrt{g}^w.$ The expression therefore equals
\begin{eqnarray}
&&\sqrt{g}^w\left( \left( \sqrt{g}^{-w}\wp ^\mu \right) _{,\rho }+\left( 
\sqrt{g}^{-w}\wp ^{\lambda }\Gamma _{\lambda \rho }^{\mu }\right) \right) 
\\
&=&\wp _{,\rho }^\mu +\wp ^\lambda \Gamma _{\lambda \rho }^\mu -w\frac{\sqrt{%
g}_{,\rho }}{\sqrt{g}}\wp ^\mu .  \nonumber
\end{eqnarray}
This argument can be extended to give the covariant derivative of an
arbitrary rank tensor of arbitrary weight $w$,
\begin{eqnarray}
T_{\beta_1\beta_2..;\rho}^{\alpha_1\alpha_2..} 
&=&T_{\beta_1\beta_2..,\rho}^{\alpha_1\alpha_2..}+T_{\beta_1\beta_2..}
^{\mu \alpha_2..}\Gamma _{\mu \rho }^{\alpha_1}+... \\
&&-T_{\nu \beta_2..}^{\alpha_1\alpha_2..}\Gamma _{\beta_1\rho }^{\nu }-...-w%
\frac{\sqrt{g}_{,\rho 
}}{\sqrt{g}}T^{\alpha_1\alpha_2..}_{\beta_1\beta_2..}.\medskip   \nonumber 
\end{eqnarray}
\subsection{Integrals of Second Rank Tensors}
\qquad Second rank tensors are most easily handled if they are
antisymmetric. Consider an antisymmetric second rank tensor $F^{\alpha
\beta }$. We can take the following covariant derivative: 
\begin{equation}
F_{;\beta }^{\alpha \beta }=F_{,\beta }^{\alpha \beta }+F^{\alpha \mu
}\Gamma _{\mu \beta }^\beta +F^{\mu \beta }\Gamma _{\mu \beta }^\alpha .
\end{equation}
Since $F^{\alpha \beta }$ is antisymmetric and the Christoffel symbols are
symmetric the $F^{\mu \beta }\Gamma _{\mu \beta }^\alpha $ term vanishes
leaving: 
\begin{equation}
F_{;\beta }^{\alpha \beta }=F_{,\beta }^{\alpha \beta }+F^{\alpha \mu
}\Gamma _{\mu \beta }^\beta .
\end{equation}
As before, we write $\Gamma _{\mu \beta }^\beta =\frac{\left( \sqrt{g}%
\right) _{,\mu }}{\sqrt{g}}$ giving (after relabelling dummy indices) 
\begin{equation}
F_{;\beta }^{\alpha \beta }=F_{,\beta }^{\alpha \beta }+F^{\alpha \beta }%
\frac{\left( \sqrt{g}\right) _{,\beta }}{\sqrt{g}}.
\end{equation}
Therefore, similar to the vector case 
\begin{equation}
F_{;\beta }^{\alpha \beta }=\frac 1{\sqrt{g}}\left( \sqrt{g}F^{\alpha
\beta }\right) _{,\beta }.
\end{equation}
\subsection{Killing Vectors} 
We can exploit symmetries in the space-time to aid
us in integration of second rank tensors. For example, does the metric
change at all under a translation from the point $x=x^{\mu }$ to $\widetilde{%
x}= x^{\mu }+\epsilon k^{\mu }(x)?$ This change is measured by the \textit{%
Lie derivative }of the metric along $k$, 
\begin{equation}
\pounds _kg_{\mu \nu }=\lim_{\epsilon\rightarrow 0}\frac{%
g_{\mu \nu }(x)-g_{\mu \nu }(\widetilde{x)}}{\epsilon }.
\end{equation}
If the metric does not change under transport in the $k$ direction then the
Lie derivative vanishes. This condition implies the Killing equation 
\begin{equation}
k_{\nu ;\mu }+k_{\mu ;\nu }=0.
\end{equation}
The solutions (if any) to this equation are called \textit{killing vectors}%
.\newline 
\newline
\textit{\textbf{Time-like Killing vector of a spherically symmetric 
space-time}}:\medskip 

Consider the following line element: 
\begin{equation}
ds^2=-\alpha ^2dt^2+a^2dr^2+r^2d\theta ^2+r^2\sin ^2(\theta )d\phi ^2
\end{equation}
where $\alpha $ and $a$ are functions of the coordinates. The metric will be 
\textit{stationary }if the metric is time independent in some coordinate
system.. That is, 
\begin{equation}
\frac{\partial g^{\mu \nu }}{\partial x^0}=0.
\end{equation}
Where $x^0$ is a time-like coordinate. We write out the full expression for
the Lie derivative of the metric 
\begin{equation}
\pounds _kg_{\mu \nu }=k^\gamma g_{\mu \nu ,\gamma }+g_{\mu \gamma }k_{,\nu
}^\gamma +g_{\nu \gamma }k_{,\mu }^\gamma .
\end{equation}
Setting this equal to zero, a time like solution satisfying this equation is
the vector field 
\begin{equation}
k^\alpha =\delta _0^\alpha .
\end{equation}
Substituting this into (30) we get 
\begin{equation}
\pounds _kg_{\mu \nu }=\delta _0^\gamma g_{\mu \nu ,\gamma }
\end{equation}
which equals zero from (29). Therefore $\delta _0^\alpha $ is a killing
vector field for the stationary spherically symmetric space-time.$%
\blacksquare \medskip $

\qquad Consider the conservation law 
\begin{equation}
T_{;\nu }^{\mu \nu }=0
\end{equation}
Where $T$ is the stress energy tensor. We cannot integrate over this as we
did in the previous section since we would not be integrating over a scalar
(due to the presence of a free index). Therefore in general there is no
Gauss' law for tensor fields of rank two or higher. If we can find a killing
vector field in the space we can use the Killing equation to form the
following equation: 
\begin{equation}
\left( k_\mu T^{\mu \nu }\right) _{;\nu }=k_{\mu ;\nu }T^{\mu \nu }+k_\mu
T_{;\nu }^{\mu \nu }=0
\end{equation}
(note that the second term equals zero from (27) and therefore the first
term equals zero as well). We then proceed as follows: 
\begin{equation}
(k_\mu T^{\mu \nu })_{;\nu }=0=J_{;\nu }^\nu 
\end{equation}
to which we can apply Gauss' law as before. 
\begin{equation}
\int \left( \sqrt{g}J^\nu \right) _{,\nu }d^4x=\oint J^\nu \widehat{n}_\nu 
\sqrt{g}d^3x\medskip 
\end{equation}
\textit{\textbf{The Energy of a Scalar Field:\medskip}}

\qquad The Einstein field equations can be written in mixed form as 
\begin{equation}
8\pi \left( T_\nu ^\mu -\frac 12\delta _\nu ^\mu T\right) =R_\nu ^\mu .
\end{equation}
If we choose a time-like killing vector $k_{(t)}=\frac \partial {\partial t}$%
, we can form a mass integral of the form 
\begin{equation}
M=-\frac 1{16\pi }\int \left( T_\nu ^\mu -\delta _\nu ^\mu T\right)
k_{(t)}^\nu dS_\mu .
\end{equation}
Where $dS_\mu =\widehat{n}_\mu\sqrt{g} d^3x$ has the following components: 
\begin{equation}
\widehat{n}_\mu d^3x=\left(
dx^1dx^2dx^3,dx^0dx^2dx^3,dx^0dx^1dx^3,dx^0dx^1dx^2\right) .
\end{equation}
Equation (38) can be integrated to give the energy of the scalar field by
noting that the scalar field has the same stress-energy tensor as a
pressure=density perfect fluid. The stress-energy tensor of the real scalar
field can also be written as 
\begin{equation}
T^{\alpha \beta }=\left( 1/4\pi \right) \left( \phi ^{;\alpha }\phi ^{;\beta
}-\frac 12g^{\alpha \beta }\phi _{;\gamma }\phi ^{;\gamma }\right) .
\end{equation}

\end{document}